\documentclass{vgtc}                          

\graphicspath{{figures/}{pictures/}{images/}{./}} 

\usepackage{times}                     

\usepackage{tabu}                      
\usepackage{booktabs}                  
\usepackage{lipsum}                    
\usepackage{mwe}                       

\usepackage{mathptmx}                  

\usepackage{graphicx}
\usepackage{subcaption}

\onlineid{0}

\vgtccategory{Research}

\vgtcinsertpkg

\title{{LegiScout: A Visual Tool for Understanding  Complex Legislation}}

\author{Aadarsh Rajiv Patel\thanks{e-mail: aadarshrajiv.patel@stonybrook.edu}\\ %
        \scriptsize Stony Brook University %
\and Klaus Mueller\thanks{e-mail: mueller@cs.stonybrook.edu}\\ %
        \scriptsize Stony Brook University}

\teaser{
  \centering
  \includegraphics[width=\linewidth]{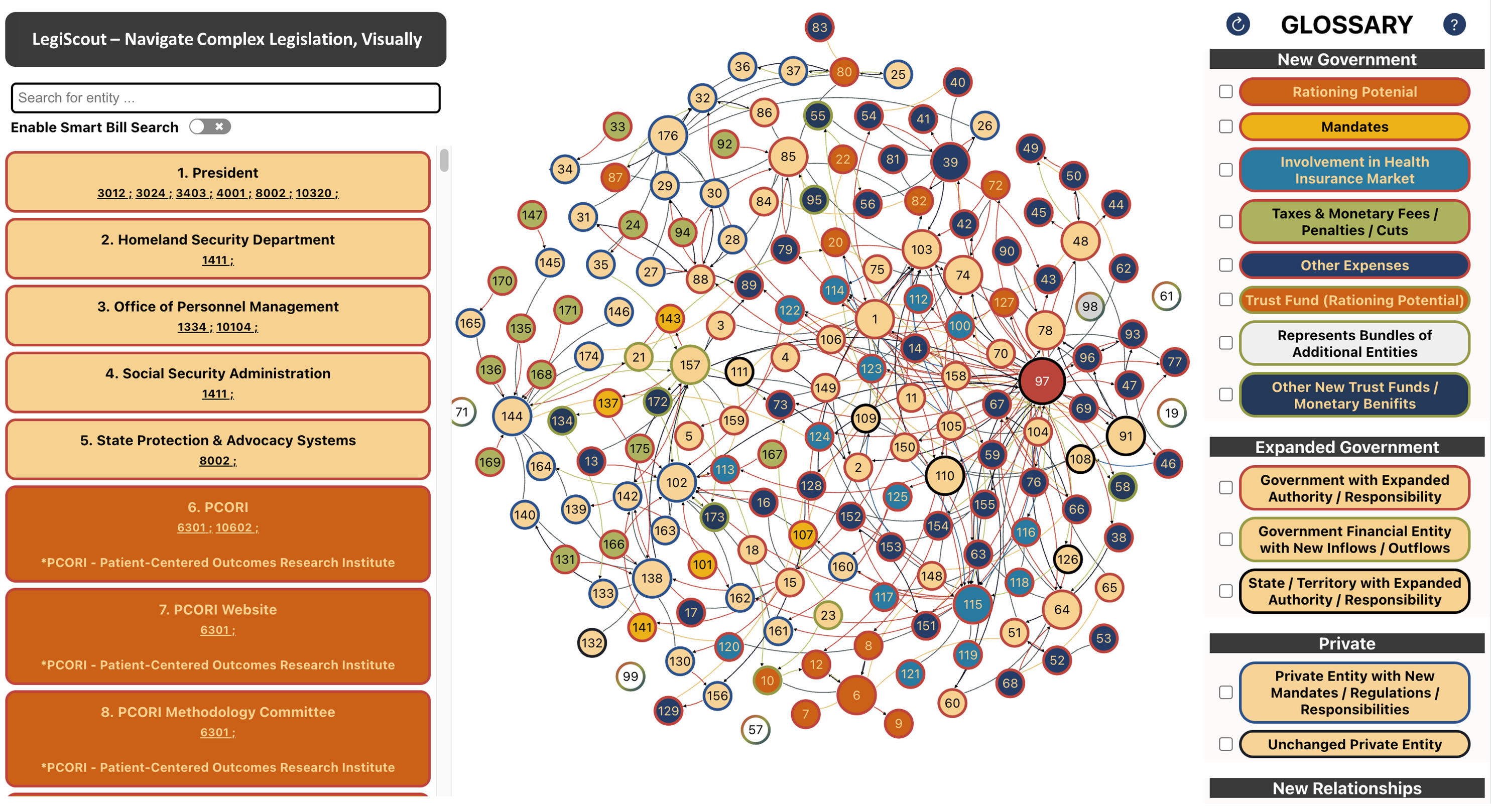} 
  \caption{LegiScout dashboard. The central panel displays the interactive, force-directed \textit{legislative-organizational graph (LOG)}, where users can click on nodes to highlight immediate relationships. On the left, a scrollable list of entities provides quick access to major organizational actors within the legislation; selecting one centers the graph on its connections. On the right, a glossary offers filterable categories describing the roles and attributes of each node. At the top left, a search bar enables users to enter legislative terms—such as "appropriation" or "mandate"—and highlights relevant nodes through BERT-assisted semantic linking.}
  \label{fig:teaser}
}

\abstract{
Modern legislative frameworks, such as the Affordable Care Act (ACA), often involve complex webs of agencies, mandates, and interdependencies. Government-issued charts attempt to depict these structures but are typically static, dense, and difficult to interpret—even for experts. We introduce LegiScout, an interactive visualization system that transforms static policy diagrams into dynamic, force-directed graphs, enhancing comprehension while preserving essential relationships. By integrating data extraction, natural language processing, and computer vision techniques, LegiScout supports deeper exploration of not only the ACA but also a wide range of legislative and regulatory frameworks. Our approach enables stakeholders—policymakers, analysts, and the public—to navigate and understand the complexity inherent in modern law.
} 


\begin{document}



\maketitle

\section{Introduction}
Modern legislation often creates complex systems involving federal and state agencies, regulatory bodies, private actors, and funding mechanisms. These entities are interwoven through mandates, oversight, and interagency dependencies, forming policy ecosystems that are difficult to interpret \cite{rittel1973dilemmas}. Traditional visualizations typically rely on dense, static layouts that obscure key relationships and overwhelm users, particularly without interactivity \cite{heer2012interactive, munzner2014visualization, yi2007toward}. This poses a significant challenge for stakeholders trying to navigate the structural complexity of large-scale legislation.


A prominent example of policy complexity visualized at scale is the “Health Care Chart” produced in 2010 by the U.S. Joint Economic Committee (Republicans) \cite{jec2012obamacare}, intended to illustrate the organizational structure of the Affordable Care Act (ACA, see Figure 2). The chart’s dense, static layout depicts hundreds of entities and connections, visually emphasizing the law’s scale but offering little interpretive utility. While rhetorically powerful, it exemplifies the cognitive overload that static visualizations can produce when faced with legislative systems of this magnitude. Our work reimagines such diagrams—not by simplifying the policy landscape, but by making its structure explorable, searchable, and intelligible through interactive visualization.

\begin{figure*}[ht]
  \centering
    \includegraphics[width=0.8\linewidth]
    {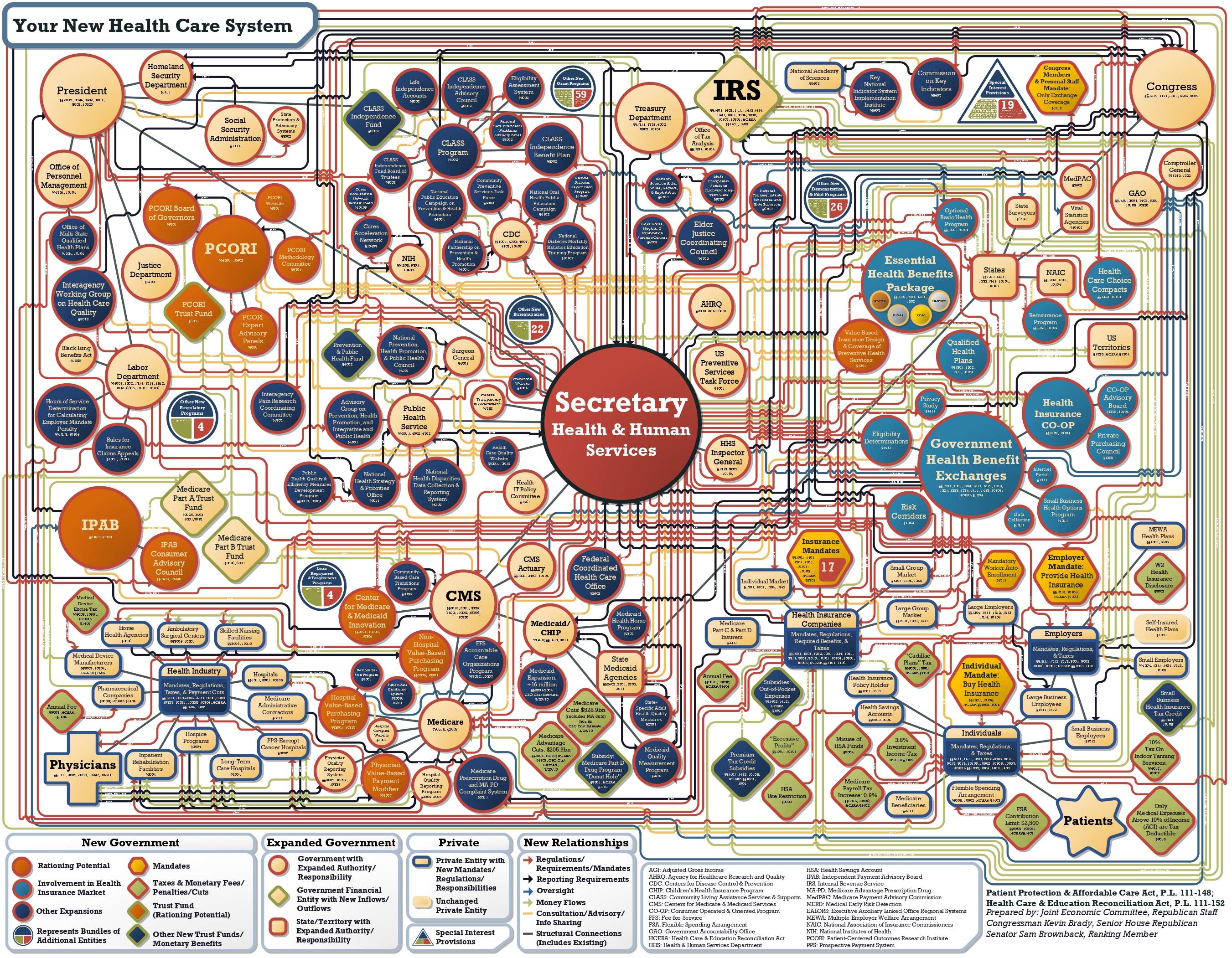}
    \label{fig:acaVis}
    \caption{The “Health Care Chart” produced by the Joint Economic Committee Republicans (2010) \cite{jec2012obamacare}. This static diagram depicts the organizational complexity of the Affordable Care Act (ACA) through hundreds of entities and interconnections. Originally created to highlight the scale of the legislation, it underscores the challenges of interpreting dense policy structures through traditional visualizations.}
    \vspace{-10 pt}
\end{figure*}

The motivation for our work is hence twofold: to improve the accessibility of complex policy frameworks through intuitive, interactive visualizations, and to develop a generalizable methodology that applies across large-scale legislative and bureaucratic systems \cite{borner2003visualizing}. We present LegiScout, a system that transforms dense, static policy structures into explorable, force-directed graphs. While demonstrated using the Affordable Care Act (ACA), LegiScout addresses broader challenges found in systems such as the U.S. Tax Code, the Dodd–Frank Act, infrastructure legislation, environmental regulations, and immigration law. It is designed for a wide range of users—including policymakers, analysts, journalists, educators, and advocates—who need to interpret, communicate, or explore institutional complexity. By reimagining how these multifaceted architectures are visualized, LegiScout contributes to the growing field of policy visualization and supports transparency, accountability, and public engagement \cite{kohlhammer2012toward, desouza2017big}.

\section{Related Work}
Our approach draws from several areas of prior research, including interactive visualization, knowledge graph construction, and policy informatics. Organizational and legislative structures are often communicated through static charts, which become difficult to parse as complexity increases. In network visualization, force-directed layouts have long been used to reveal structure in dense graphs \cite{fruchterman1991graph}, offering an intuitive way to explore connections between entities. These techniques have since been extended in the information visualization community to support interactivity and scalability \cite{shneiderman1996eyes, borner2003visualizing}. We build on this foundation to apply such methods to the legislative domain, where interactive graph-based representations remain underexplored.

In parallel, advances in natural language processing (NLP) and knowledge graph construction have enabled the extraction of structured information from unstructured legislative text. Methods such as BERT-based semantic linking \cite{zhang2019ernie} facilitate connections between user queries and graph elements, enhancing the usability of complex legal documents. Recent work on legal text understanding has also demonstrated the feasibility of bridging legislative language with structured representations \cite{chalkidis2021lexglue}. From a broader perspective, policy informatics research has emphasized the importance of data-driven tools for improving governance and public sector decision-making \cite{desouza2017big}. Our work builds on these insights to support stakeholders—ranging from policy analysts to the general public—in interpreting and navigating complex legislative ecosystems through interactive and intelligent visualization.

\section{Methodology}


To guide the design of our system (see Figure 1), we identified core requirements based on the complexity of legislative structures and expert feedback. These include: (1) automated extraction of entities and their interrelationships from legislative and organizational documents; (2) rendering these structures as interactive, force-directed graphs; (3) enabling semantic search that links key terms—such as "appropriation" or "funding"—to graph elements; (4) ensuring scalability across various legislative texts, including large bills and reconciliation packages; 
and (6) accommodating input from static organizational charts, such as government-issued PDF documents. In the following we describe each of these steps in detail, using the ACA as an example.


\subsection{Data Extraction and Processing}

\textbf{Optical Character Recognition (OCR):} We used Tesseract to extract text from the chart image, applying pre-processing techniques such as binarization and noise reduction \cite{smith2007overview}. While OCR captured many textual elements, it struggled with overlapping text, non-standard layouts, and lacked awareness of structural relationships, necessitating further methods.

\textbf{Computer Vision-Based Shape Detection:} We used standard techniques like the Hough Transform \cite{duda1972use} and Contour Detection \cite{suzuki1985topological} to identify shapes and connections in the chart, with post-processing by morphological operations \cite{gonzalez2002digital} to improve consistency. However, challenges such as line thickness variation and element overlap reduced effectiveness.

\textbf{Graph-Based Image Processing:} To cluster related elements, we used edge-detection and morphological transformations within a graph-based segmentation framework. This approach showed promise but was limited by layout complexity, particularly in dense regions where connections were misclassified \cite{noack2009energy}.

\textbf{Manual Data Extraction:} Given the limitations of automated methods, we semi-manually curated a complete dataset of nodes and edges from the chart, including metadata such as entity types (e.g., federal agencies, insurers) and relationship types (e.g., regulatory, funding). 

\subsection{Rendering of the Graph}

\textbf{Graph Layout:} We refer to our graphical structure as a \textit{legislative-organizational graph (LOG)}—a directed, heterogeneous network that captures how laws instantiate agencies, programs, and inter-agency relationships. Entities are mapped as nodes and legal or administrative connections as directed edges. To visualize the LOG, we employ a force-directed layout \cite{fruchterman1991graph}, which simulates physical forces: nodes repel one another to reduce overlap, while edges act as springs that draw related nodes together. 


\textbf{Graph Elements:} Nodes represent ACA entities—such as federal agencies, insurers, and healthcare providers—enriched with metadata (e.g., name, type, and role). Edges represent relationships like regulatory oversight, funding flows, or collaborative partnerships, with associated metadata describing type and directionality. 

\textbf{Visual Encoding:} To support visual clarity, nodes and edges are styled based on their roles: e.g., federal agencies may appear as large blue circles, healthcare providers as smaller green squares. Edge styles indicate relationship types—solid lines for regulatory, dashed for funding, and dotted for partnerships—enabling users to quickly interpret entity roles and interactions. Tooltips provide on-demand details when hovering over visual elements.


\textbf{Interactivity Features:} Several features enhance user navigation and comprehension. For scalability, users can zoom and pan across the graph to explore areas of interest. A search bar enables quick location of entities, and filters support focus on policy areas such as Medicaid or private insurance. Hovering highlights a node’s direct connections, helping trace relationships across the LOG.

\textbf{Challenges and Solutions:} Implementing an effective LOG visualization posed several challenges. Overlapping node labels were resolved using collision detection and dynamic positioning. To counteract layout instability, we fine-tuned force simulation parameters and introduced a "freeze" mode to lock node positions. To address visual clutter in densely connected regions, we implemented hierarchical clustering with expand-collapse functionality. Graph rendering and interactivity are implemented using D3.js.


\subsection{Semantic Search}
In addition to keyword-based search that identifies graph entities by their names and tags, our interface also incorporates a semantic search feature that allows users to retrieve relevant bill sections based on the meaning of their queries. For instance, entering "coverage for dependents up to age 26" retrieves the exact ACA sections addressing that policy — even if the phrasing differs from the user's query. The system uses state-of-the-art sentence embeddings  \cite{reimers2019sentence, devlin2019bert} to match user input with the most semantically relevant chunks of the ACA legislation.

\begin{figure}[htbp]
  \centering

  \begin{subfigure}[t]{\linewidth}
    \includegraphics[height=0.22\textheight]{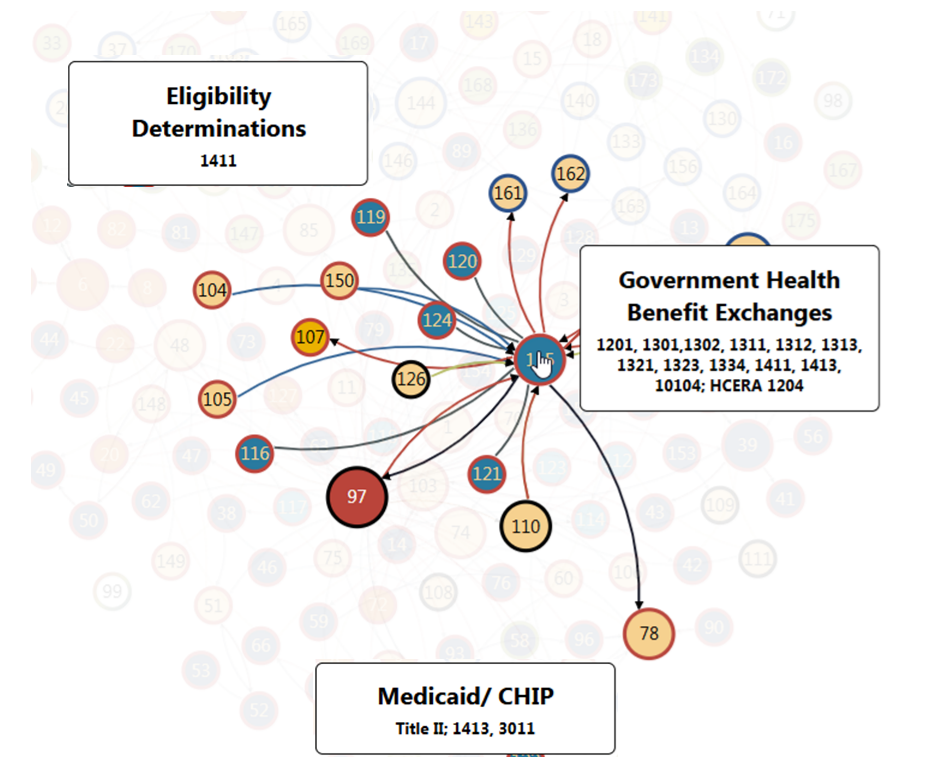}
    \caption{Government Health Benefit Exchanges at the center of exploration.}
  \end{subfigure}

  \vspace{0.5em}

  \begin{subfigure}[t]{\linewidth}
    \includegraphics[height=0.22\textheight]{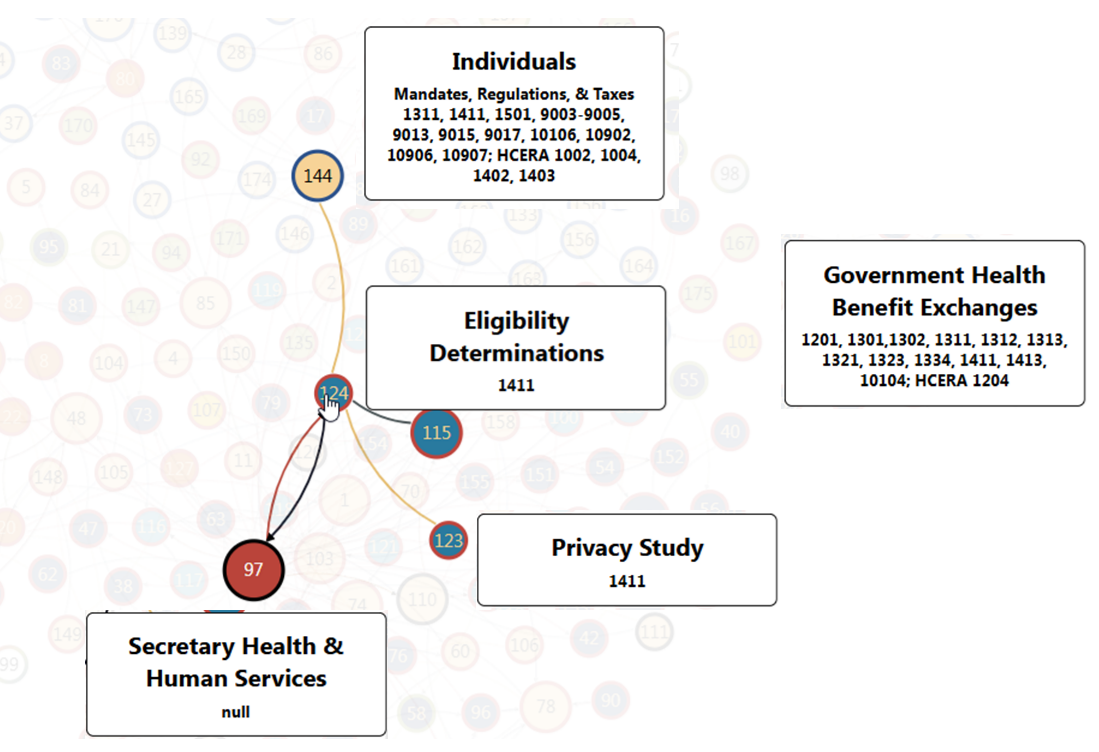}
    \caption{Eligibility Determinations reveal privacy and individual-level nodes.}
  \end{subfigure}

  \vspace{0.5em}

  \begin{subfigure}[t]{\linewidth}
    \includegraphics[height=0.22\textheight]{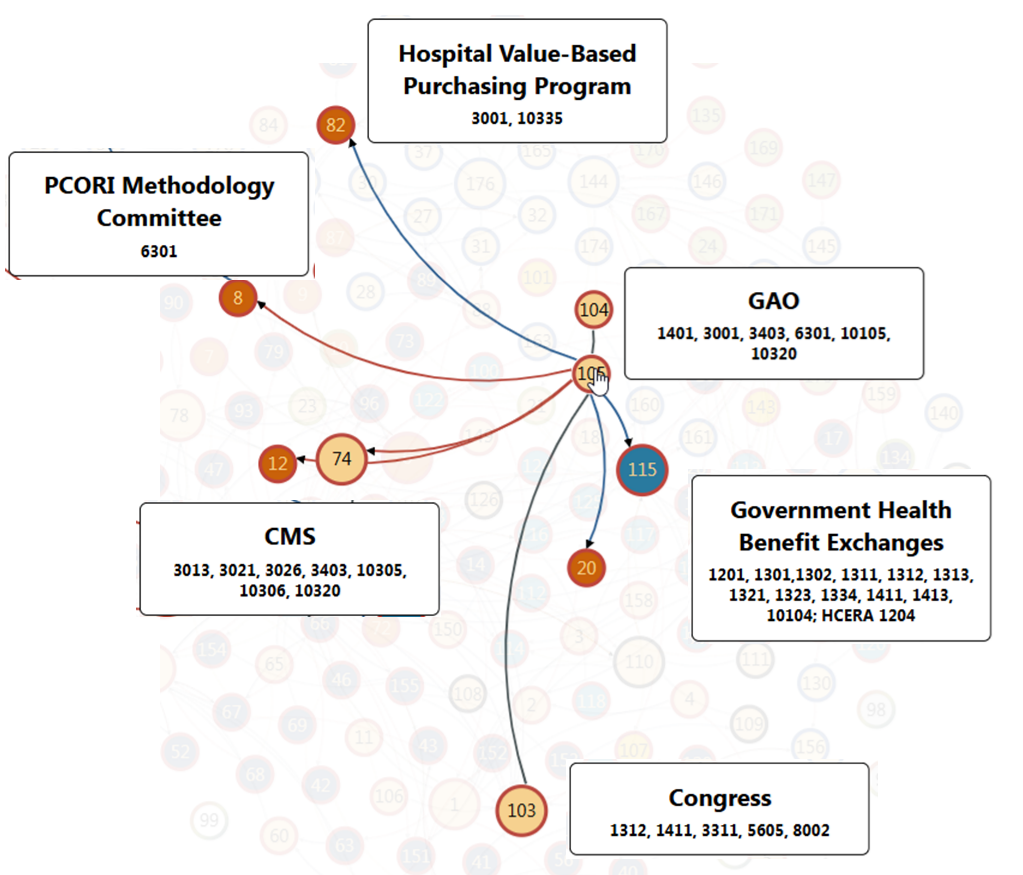}
    \caption{GAO oversight surfaces through eligibility node.}
  \end{subfigure}

  \vspace{0.5em}

  \begin{subfigure}[t]{\linewidth}
    \includegraphics[height=0.22\textheight]{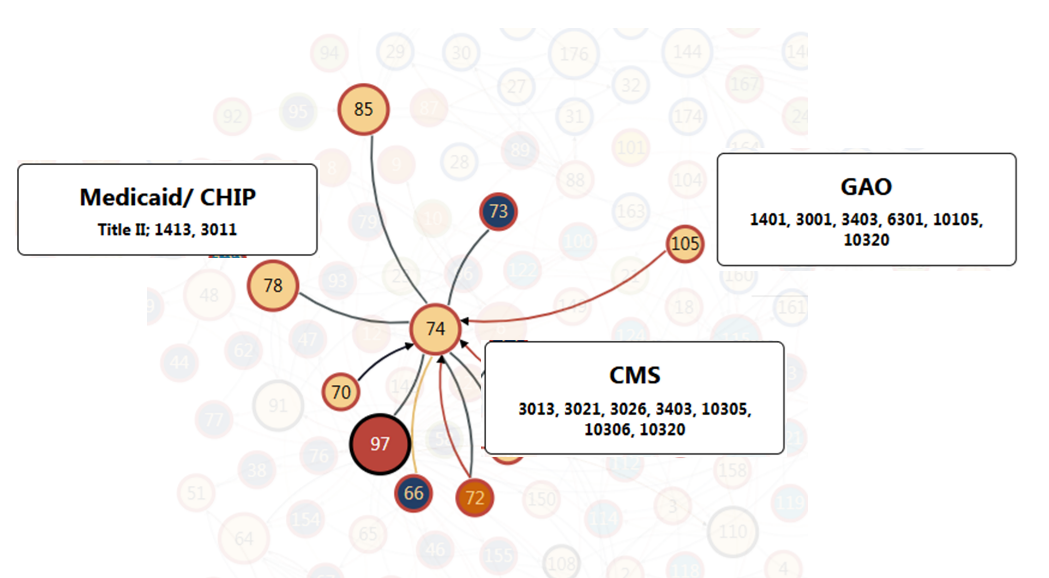}
    \caption{CMS as an implementation hub with links to exchanges and Medicaid.}
  \end{subfigure}

  \caption{Exploratory path through the ACA policy graph using LegiScout. The user begins with exchanges (a), continues to eligibility (b), discovers GAO oversight (c), and ends with CMS as a key implementer (d).}
  \label{fig:avigator-vertical}
\end{figure}

\subsection{Visual Interface and User Experience}

The graph includes several interactive features to enhance usability:

\textbf{Hover Effects}: When a user hovers over a node, its connections are highlighted, making it easier to trace relationships. For example, hovering over a node representing HHS might highlight all entities it regulates or funds. 

\textbf{Click to Lock}: Users can click on a node to lock it in place, preventing it from moving during layout adjustments. This is particularly useful when exploring densely connected graph sections. 

\textbf{Dynamic Updates}: The graph layout is dynamically updated as users interact with it-- expanding a cluster of nodes triggers a recalculation of the layout to accommodate the new elements.




\textbf{Contextual Information on Hover}: 
When a user hovers over a node or link, a tooltip displays additional contextual information. For example, hovering over a node representing an insurer displays its name, type, and the number of policies it manages. 

\textbf{Zooming and Panning}: 
The graph includes zoom and pan functionalities to facilitate navigation. Users can zoom in to explore densely connected sections of the graph or zoom out to view the entire structure. Panning allows users to move around the graph, ensuring that no part of the visualization is out of reach.




\textbf{Direct Bill Access via Clickable IDs}:
Each bill ID within the interface is clickable, allowing users to directly access the original legislative text. When a user clicks on a bill ID, the system opens a pdf file of the official legislation document at the exact page where that specific bill is discussed. This seamless integration of navigation and document access supports deeper exploration and provides analysts with immediate access to the authoritative source.

\section{Case Study}
For a tour through the system we shall follow Maria, a graduate student in public health, who is preparing a presentation on how the Affordable Care Act (ACA) expanded access to health coverage. She’s read summaries of the law, but the policy diagrams she’s seen are overwhelming—dense, static, and hard to follow. Wanting to understand how the ACA actually works in practice, she points her web browser to LegiScout to explore the law’s complex structure. The interface shown in Figure 1 comes up -- in the following description we focus on the center panel only, 

Maria begins by clicking on Government Health Benefit Exchanges, one of the law’s most visible features. On the screen, a web of connected entities comes into view (Figure 3a), including agencies like the Department of Health and Human Services, programs like Medicaid/CHIP, and references to eligibility determinations. It’s clear that the exchanges aren’t standalone—they're woven into a broader implementation network.

She clicks on the node “Eligibility Determinations” to investigate how individuals qualify for coverage. In the new view (Figure 3b), she notices two intriguing nodes: one labeled “Individuals”, and another, “Privacy Study”. That catches her attention—she hadn’t considered that the ACA included formal analysis of how personal data would be handled during enrollment. LegiScout helps her see that data governance was built into the legislation itself.

Following the thread of oversight, Maria clicks on the GAO (Government Accountability Office) node (Figure 3c). Now she sees a structure of accountability emerging: GAO is linked not only to eligibility but also to evaluations of cost-saving programs and insurance exchange performance. This part of the law isn’t about direct service—it’s about making sure those services are accountable and effective.

Finally, Maria navigates to CMS, the Centers for Medicare and Medicaid Services (Figure 3d). It quickly becomes clear just how central CMS is: it's linked to nearly every implementation branch of the ACA. As she explores its neighbors, she spots PCORI—a research institute created by the law to fund patient-centered outcomes research. It’s another surprising discovery: the ACA didn’t just build coverage systems, it also invested in understanding what care actually works best for patients.

By the end of her session, Maria has followed a path from high-level coverage policy to implementation details, oversight bodies, data privacy, and research infrastructure. What began as a dense, bureaucratic chart has turned into a story of how policy becomes practice—and how LegiScout  makes that story accessible.


\section{User Study and Expert Feedback}
{To evaluate the utility and usability of our system, we conducted an informal user study with four senior and junior analysts from a renowned federal grant and government relations firm based in Washington, D.C. The participants regularly engage with complex legislative and organizational structures as part of their work and provided valuable domain-specific feedback.}

{A key concern raised was the limitation of the initial search functionality, which only allowed users to locate entities already labeled within the graph. Analysts emphasized the importance of being able to search for specific terms in the legislative text itself—such as "appropriation" and "funding"—to trace their connections to structural elements in the graph. In response, we implemented the BERT-assisted semantic search feature (see Section X) that links relevant portions of the legislative text directly to the corresponding nodes and edges in the graph.}

{Participants also highlighted the broader applicability of the system beyond the Affordable Care Act. They noted that the tool would be especially valuable for navigating large and complex bills, such as the Infrastructure Investment and Jobs Act and annual congressional appropriations bills, as well as reconciliation bills. Some even suggested that the system could benefit the general public and be extended to static organizational charts—such as those used by the Department of Defense or interdepartmental agencies—where structural complexity often hinders understanding.}

\section{Discussion and Future Work}
While our semi-manual extraction yielded a functional dataset, future work will focus on fully automating the process. Specifically, we would like to study deep learning methods, such as CNNs for object detection and GNNs for relationship inference, that have successfully been used to identify structural elements in similar bureaucratic charts \cite{he2017mask}.  
At the same time, NLP techniques can support semantic categorization and link entities to legislative text \cite{smith2007survey}.

On the visualization side, we wish to further improve usability and layout clarity. Reinforcement learning or physics-based simulations may enhance graph readability, while features like draggable nodes, filtering, and personalized views could support deeper exploration. Finally, connecting the system to live policy data would keep it current and actionable.

Looking further down the road, we aim to generalize this approach to other major laws and static organizational charts. By combining computer vision, NLP, and user-centered design, we envision a scalable pipeline for generating interactive legislative-organizational graphs across a broad range of policy domains.

\section{Conclusions}
We believe that this project demonstrates the feasibility of simplifying complex bureaucratic charts through structured, interactive visualization. While semi-manual curation was key, future work combining computer vision, NLP, and graph analytics could enable scalable, automated solutions. The approach is applicable beyond the ACA, offering a blueprint for clarifying other legislative or organizational systems. Continued improvements in interaction design will further enhance accessibility for diverse users.


\section*{Supplemental Materials}
\label{sec:supplemental_materials}
A demo video has been submitted along with the paper. More supplemental materials, including datasets, code, and additional visualizations, are available on \href{https://github.com/aadarsh0301/aca_vis/tree/dev}{GitHub}. The repository is released under a CC BY 4.0 license. An app of the system is available on \href{https://aca-vis-ee7ab363c3c5.herokuapp.com/#8002}{Heroku}.




\bibliographystyle{abbrv-doi}
\bibliography{template}
\end{document}